\documentclass[12pt]{article}
\newtheorem{theorem}{Theorem}
\newcommand{\alg}[1]{\mathcal{#1}}

\begin{document}

\begin{center}
\Huge{Hyperentangled States}
\end{center}
\begin{center}
\Large{Rob Clifton}
\end{center}
\begin{quote}
 \emph{Department of Philosophy, 1001 Cathedral of
Learning, University of
Pittsburgh, Pittsburgh, PA\ 15260 (rclifton+@pitt.edu).}
\end{quote}
\begin{center}
 \Large{David V. Feldman}
 \end{center}
 \begin{quote}
\emph{Department  of  Mathematics and  Statistics,
University of New Hampshire, Durham, NH\ 03824 (dvf@christa.unh.edu).}
\end{quote}
\begin{center}
\Large{Michael L. G. Redhead}
\end{center}
\begin{quote}
\emph{Department of History and Philosophy of Science, Free School
Lane, Cambridge, England\ CB2 3RH (mlr1000@cam.ac.uk).}
\end{quote}
\begin{center}
 \Large{and}
 \end{center}
 \begin{center}
 \Large{Alexander Wilce}
 \end{center}
 \begin{quote}
\emph{Department  of  Mathematics and  Computer
Science, Juniata  College, Huntingdon, PA\ 16652 (wilce@juniata.edu).}
\end{quote}
\begin{center}
 \Large{November 1997}
 \end{center}

\pagebreak
\begin{center}
\Large{Abstract}
\end{center}
\begin{quote}
We investigate a new class of entangled states, which we call
\emph{hyperentangled}, that have EPR correlations identical to those
in the vacuum state of a relativistic quantum field.  We show that
whenever hyperentangled states exist in any quantum theory, they are
dense in its state space.  We also give prescriptions for constructing
hyperentangled states that involve an arbitrarily large collection of
systems.

\end{quote}
\pagebreak

\section{EPR Correlations in the Vacuum}

It is well-known that the Bell
inequality is generically violated in the states
of free relativistic quantum
fields (see \cite{but} for a recent review).  This
reinforces the point, usually driven home in the context of
nonrelativistic quantum mechanics, that quantum correlations cannot be
explained locally in terms of common causes or pre-existing elements
of reality.   However, it is important to remember that
 Bell's original derivation of his inequality \cite{bell} did not simply
assume local
realism from the outset.  Rather, Bell (pp. 14-5) launched his derivation
from a
version of the Einstein-Podolsky-Rosen (EPR) argument \cite{EPR}
in order to motivate his assumption that elements of reality exist
for all the observables that figure in the inequality he derives.
Granting that EPR had
successfully established the dilemma that either locality must fail
or quantum mechanics fails to account for certain
elements of reality, Bell showed that even if we grasp the
second horn of EPR's dilemma, locality must fail \emph{in any case}
on pain of failing to account for the correlations actually obtained.
This raises the question of whether one can reproduce Bell's
own EPR-inspired argument for nonlocality in relativistic
quantum field theory by employing, for example, vacuum
correlations.

 Recall that to apply their
sufficient condition for identifying an element of reality,
EPR needed to exploit correlations
with the feature that if a local
measurement of a particle's position (or momentum) were performed,
a distant particle's position
(or momentum) could be predicted with certainty.  So EPR's
argument relies essentially on the availability of \emph{maximal}
correlations.  Are such
correlations present in the vacuum?  Indeed they are, all over the
place. Redhead \cite{red1,red2} has recently
 shown that
 any field state with bounded energy---the vacuum state in
 particular---dictates a host of maximal correlations
 between the values of observables associated with \emph{any} two
 spacelike-separated regions.  Thus vacuum correlations supply plenty
 of resources to run Bell's EPR-inspired derivation of his inequality.

 On the other hand, even if the Bell inequality were not violated in this
context, so
 allowing the possibility of a common cause explanation of the
 correlations by events in the overlapping backwards lightcones of the
 correlated events, this would still not give an acceptable local
 explanation of the correlations.  This is because the vacuum is
 time-translation invariant, so any common cause events employed to
 explain correlations at a later time are inevitably involved in
 correlations that themselves need explaining.   It seems we
 must either embark on
 an infinite regress of explanation, or accept the correlations
 as `brute facts'.

 Of course the situation is not so hopeless as that.
 Even if no satisfactory \emph{causal} explanation of maximal vacuum
correlations is
 available, this does not preclude
 pursuing a deeper understanding of those correlations
 in purely quantum-mechanical terms.
 The present paper aims to do precisely this, following on from the initial
 investigations undertaken in \cite{red1,red2,wag}.
 In particular, we uncover necessary and sufficient
 conditions for maximal correlations like those in the vacuum
 to obtain in the states posited by \emph{any} quantum theory
 (not necessarily relativistic
 or field-theoretic), and investigate how generic such states are.
 But before we can get started, we need to delve into Redhead's
 argument \cite{red1} for the existence of maximal vacuum correlations.

 \section{Maximal Vacuum Correlations}

 The formal proof
 of the pervasiveness of maximal vacuum correlations
  proceeds within the algebraic approach to relativistic quantum
  field theory \cite{haag, horuzy}.
  On that approach,
one associates with each bounded open region $O$ in Minkowski
spacetime $M$ a von Neumann algebra $\alg{U}(O)$ of bounded operators
whose self-adjoint
elements represent
observables measurable in region $O$ (constructed from field operators smeared
with test functions having support in $O$).  Each algebra in the
collection $\{\alg{U}(O)\}_{O\subseteq M}$ acts on the same Hilbert space
 $H$ consisting of states of the
entire field on $M$.

If these algebras satisfy certain plausible general
conditions---notably that algebras associated with
spacelike-separated regions must commute and that the energy-momentum
spectrum of the field must be confined to the forward lightcone (see
\cite{but} for
the complete list of conditions)---then it becomes possible to
prove the Reeh-Schlieder theorem \cite{RS}.  It is this theorem that is
pivotal for
establishing maximal correlations in the vacuum.
If we call a field state $\Psi$ \textbf{cyclic for} $\alg{U}(O)$ 
whenever
the set $\{A\Psi: A\in \alg{U}(O)\}$ is dense
in $H$, then the Reeh-Schlieder theorem asserts that
for \emph{any} bounded open region $O$, the
vacuum state $\Omega$ (or any other field state $\Psi$ with bounded energy) is
cyclic for $\alg{U}(O)$.

In addition to the Reeh-Schlieder theorem, Redhead's argument uses
a simple result about commuting von Neumann algebras $\alg{U}(O)$
and $\alg{U}(O')$, viz. that if a vector $\Psi$ is cyclic for one of the
algebras $\alg{U}(O)$
or $\alg{U}(O')$ it must separating for the other, where $\Psi$ is called a
\textbf{seperating vector for} an algebra if no nonzero operator
in the algebra maps $\Psi$ to $0$.
Since the proof of this result is elementary, and we will revisit
 it a number of times below, we give its proof.
If, say, $\Psi$ is cyclic for $\alg{U}(O)$ then there is
a sequence of operators $\{A_{m}\}\subseteq\alg{U}(O)$ that commute
with any $A'\in \alg{U}(O')$
and can be
applied to $\Psi$ to approximate any vector $\Phi\in H$ as closely as
desired.  Therefore, if $A'\Psi =0$ we have
\begin{equation}
 A'\Phi = A'\lim A_{m}\Psi =\lim A_{m}A'\Psi = 0
 \end{equation}
for any $\Phi \in H$, which implies $A'=0$.

Now
call a von Neumann algebra $\alg{U}(O')$ \textbf{maximally correlated with}
a von Neumann algebra $\alg{U}(O)$ in the state $\Psi$ if for any nonzero
projection operator
$P'\in\alg{U}(O')$
and any $\epsilon>0$, there is a nonzero projection operator $P\in\alg{U}(O)$
such that
\begin{equation}
\label{eq:Sch}
\mbox{Prob}_{\Psi}(P'=1/P=1)=1-\epsilon.
\end{equation}
 (Note that this relation between
algebras given a state $\Psi$ need not be symmetric.)  What Redhead's
argument
establishes is that if any state $\Psi$ is cyclic for $\alg{U}(O)$ and $O'$
spacelike-related to $O$---so that $\alg{U}(O')$ commutes with
$\alg{U}(O)$---then $\alg{U}(O')$ is maximally correlated
with $\alg{U}(O)$ in state $\Psi$.  Without pursuing the details of the
argument, the basic idea is intuitive.
Since $\Psi$
is assumed cyclic for $\alg{U}(O)$ it is
seperating for $\alg{U}(O')$, so that for any  nonzero projection
$P'\in\alg{U}(O')$ the state $\Phi = P'\Psi/\|P'\Psi\|$ is
well-defined.  Moreover, by construction of $\Phi$,
Prob$_{\Phi}(P'=1)=1$.  One then uses the possibility of getting
arbitrarily close to $\Phi$ by acting on $\Psi$ with operators
in $\alg{U}(O)$ to infer (via the spectral theorem)
the existence of a sequence of projections
$\{P_{m}\}\subseteq \alg{U}(O)$ such that $\lim$ Prob$_{\Psi}(P'=1/P_{m}=1)=1$.
Of course, what makes Redhead's theorem relevant for the physics of the
\emph{vacuum} is the Reeh-Schlieder theorem.  For together
these theorems imply that the vacuum state
 $\Omega$ is actually
 filled with
maximal correlations:
for any two spacelike-separated (bounded, open) regions
$O$ and $O'$, $\alg{U}(O')$ is maximally correlated with $\alg{U}(O)$ in the
state $\Omega$.

Having made precise the sense in which there are maximal vacuum
correlations, it may now appear that there are some obstacles to running
the EPR
argument in this setting.
In the first
place, EPR's sufficient condition for identifying an element of
reality only permits the inference to an element of reality in
situations where the outcome of measuring it can be predicted with
\emph{absolute} certainty.  For maximal vacuum correlations, all we have is
prediction
with \emph{arbitrarily high} certainty.  Is it possible to remove
this restriction?  No.
Suppose that for some nonzero  $P'\in\alg{U}(O')$ there
is a  nonzero $P\in\alg{U}(O)$ such that
Prob$_{\Omega}(P'=1/P=1)=1$.  Then Prob$_{\Omega}(P'=0/P=1)=0$ which
requires $(I-P')P\Omega=0$.  But $(I-P')P$ is an element of the
algebra generated by $\alg{U}(O)\cup\alg{U}(O')$, which is the local
algebra associated with the bounded open region $O\cup O'$ (cf.
\cite{haag}, p. 107).  Since $\Omega$ is seperating for this algebra,
it follows that
$(I-P')P=0$.  And this contradicts a well-known consequence of
 the axioms of algebraic relativistic quantum field
theory (\cite{horuzy}, p. 76): that nonzero operators
(such as $(I-P')$
and $P$) associated
with (generic) space-like separated commuting algebras
must have a nonzero product.  The upshot is
 that vacuum correlations cannot supply conditional
 predictions with absolute
certainty.  However, we see no reason why the
ability to predict the outcome of
measuring some local observable with arbitrarily high certainty
should not give one just as strong
grounds to infer the existence of an element of reality corresponding
to that observable as one would get if its value were predictable
with absolute certainty.

There also appears to be a second worry. The correlations EPR originally
exploited had the additional feature that \emph{whatever} the result of the
local
measurement of position (or momentum), the distant particle's position
(or momentum) could be predicted with certainty.  By contrast,
even if the probability for $P'=1$
given $P=1$ is close to 1 as per Eqn. \ref{eq:Sch}, it does not follow
that either
$P'=1$ or $P'=0$ must be assigned a probability near to 1 given that $P=0$ (though this may be
true for certain special choices of $P'$).
Again, however, we do not think this undermines the validity of EPR's
argument in the vacuum.  It was important to their argument that if
\emph{without in any
way disturbing a system} we can predict with certainty the outcome of
measuring one of its observables, then there exists an element of
reality corresponding to the observable.  Behind this was the idea
that one should not be able to make sharp the value of a local observable
by performing measurements at a distance.  But to the extent that
this locality assumption is plausible, it should also be plausible
to assume that one cannot make sharp the value of a local observable
by performing a distant measurement \emph{and getting some particular
outcome}.  Indeed, since Redhead's theorem tells us that for any local
observable and any possible outcome of measuring it, it is always
possible to perform a distant measurement on a second observable
and get an outcome that
dictates with (virtual) certainty the outcome of measuring the first
observable, EPR-type reasoning leads immediately to the conclusion
that there are pre-existing elements of reality in the vacuum state for
\emph{all} local observables!  Of course, while valid we are not saying
EPR's reasoning
is sound, since it is based upon locality assumptions which, in turn,
lead to the Bell inequality.

One last point before we probe deeper into the origin of maximal
vacuum correlations. These correlations are of exactly the sort
that troubled Schr\"{o}dinger \cite{Schr} when
he wrote of the `sinister'
possibility in quantum mechanics of steering a distant system into
any desired state by a suitable local measurement.
But one must not be misled into thinking that the correlations
lead to any
empirically detectable nonlocality, in violation of spacelike
commutativity.  In the vacuum, the outcome of measuring the
projection $P$ in Eqn. \ref{eq:Sch} is generally going to be probabilistic
and cannot
\emph{itself} be controlled, which is what would be needed to truly `steer'
a distant system close to an eigenstate of $P'$.  In fact, it turns
out that the
well-known Fredenhagen bound \cite{fred} on correlations in the vacuum
entails that for any given $P'\in\alg{U}(O')$,
the maximally correlated $P\in\alg{U}(O')$ in the vacuum state must have
a probability of occurring that
falls off exponentially with the minimum Lorentz distance between $O'$
and $O$ (see \cite{red1}, Sec. 3).  Moreover, the outcome of
measuring the projection $P$ in Eqn. \ref{eq:Sch} \emph{will be} certain to
be $1$ in the vacuum only in the uninteresting case when $P'$ is the
identity operator.
For
if Prob$_{\Omega}(P'=1/P=1)=1$ when $(\Omega,P\Omega)=1$, then
$I-P$ annihilates $\Omega$ which forces $P=I$
because $\Omega$ is a seperating vector for all the local
algebras.  This, in turn, means we must have Prob$_{\Omega}(P'=1)=1$,
which by the same reasoning forces $P'=I$ as well.

\section{Hyperentanglement}

Since the pervasiveness of maximal vacuum correlations rests on the
Reeh-Schlieder theorem, this is the natural place to look
for an explanation for
their presence.  The theorem is remarkable because if $O$ is, for example,
the neighbourhood of some particular point in $M$, how could acting
on $\Omega$ with operators in $\alg{U}(O)$ approximate an arbitrary
state of the field, in particular one which looks quite unlike the
vacuum in some remote region spacelike-separated from $O$?
The short answer, given by Haag (\cite{haag}, p. 102), is
that requiring the
energy-momentum spectrum of the field to be confined to the forward
lightcone forces the vacuum
to be a highly correlated state, and it is these correlations which
are `judiciously exploited'
to prove the Reeh-Schlieder
theorem.  However, the analyticity
arguments that go into the proof (cf. \cite{horuzy}, pp. 25-6)
shed little light on this, nor do they give any sense of
exactly what
structure a state needs to have to be saturated with
maximal correlations between spacelike-separated observables.

Our approach to this issue will be to abstract away from the context of
quantum field
theory and analyze the matter in terms of the more familiar concept of
entanglement.  We shall see that the sorts of states that give rise to
pervasive maximal
correlations
can occur in any quantum theory.  Furthermore, contrary to what
one might expect, these states need not differ from other entangled states in
their \emph{degree} of entanglement (according to the measure recently
proposed by Shimony \cite{shim}), but
rather involve a completely new \emph{kind} of entanglement.

In order to characterize vacuum correlations in terms of entanglement, consider
 the implications the Reeh-Schlieder theorem has
for only a finite number of mutually
spacelike-separated
regions $\{O_{i}\}_{i=1}^{n}$ $(n>1)$.
 Each of the algebras $\alg{U}(O_{i})$ has a representation on a
separable Hilbert space $H_{i}$ of $\dim >1$, and since these algebras
mutually commute
 the von Neumann algebra generated by $\bigcup_{i=1}^{n}\alg{U}(O_{i})$ can
be represented
on the tensor product Hilbert space
\begin{equation}
\label{eq:Hilbert}
 H= H_{1} \otimes H_{2} \otimes\cdots
\otimes H_{n}.
\end{equation}
Note that for any (proper) subset $S$ of the indices $\{1,\ldots,n\}$,
$H$ can be factored as
\begin{equation}
H = H_{S} \otimes H_{S'}
\end{equation}
where
\begin{equation}
H_{S}=\bigotimes_{i\in S}H_{i}\ \mbox{and}\ H_{S'}=
\bigotimes_{i\not\in S}H_{i},
\end{equation}
and $S'$ is the complement in $\{1,\ldots,n\}$ of $S$.  We shall also use
the letter $S$
 to refer to the \emph{algebra} represented
on $H_{S}$.  In that case, $S'$ will refer to the commutant of $S$, i.e.
the set of all
bounded operators on $H$ that commute with those in $S$, which are exactly
the operators represented on $H_{S'}$.  (Whether
$S$ denotes a subset of indices labelling the factors in $H$ or the
corresponding operator algebra will always be clear from context.)
By analogy with the previous section, call a state vector $v\in H$ {\bf
$S$-cyclic} if the set
\begin{equation}
\{(A\otimes I)v: A=\mbox{a bounded operator on}\ H_{S}, I=\mbox{the identity
on}\ H_{S'}\}
\end{equation}
 is dense in $H$.  Since operators in the algebra $S$ are associated
 with the
region of spacetime $\bigcup_{i\in S}O_{i}$ (itself bounded and open), the
Reeh-Schlieder theorem asserts that $\Omega$ is $S$-cyclic for all $S$.

 Now forget about the details of relativistic quantum field theory.  State
spaces having the
 tensor product form of $H$ above occur in \emph{all} quantum
 theories and are used to
 describe compound systems with
 $n$ components, such as $n$ spinless particles.  (So sometimes we
 shall refer to a subset of the indices $\{1,\ldots,n\}$ as a
\emph{subsystem}.)  In this more general
 context, there no longer need be a vacuum state, and we no longer
 have any Reeh-Schlieder theorem.  However, Redhead's theorem is
 still valid in the form:
 \begin{theorem}
 If a state $v\in H$ is $S$-cyclic, then $S'$ is maximally correlated
 with $S$ in state $v$.
 \end{theorem}
 By analogy with the vacuum, we want to
 investigate the entanglement properties of states with the
 following feature:
 \begin{quote}
 \emph{For any two nonoverlapping subsystems $S$ and $T$,
  the algebra $S$ is maximally correlated
 with the algebra $T$.}
 \end{quote}
 Anticipating the results of our investigation, it will be appropriate to
 call such states \textbf{hyperentangled}.

 The first thing to note is that hyperentangled states, so defined,
 are indeed entangled---indeed, they must be entangled
 with respect to \emph{any} of the possible factorizations of $H$ as
 $H = H_{S} \otimes H_{S'}$.
 For suppose that some state $v$ is both hyperentangled and a product
 state.  Then, on the one hand, for any nonzero projection $P'\in S'$ and
$\epsilon>0$, there
 must exist a nonzero projection $P\in S$ such that
  \begin{equation}
  \label{eq:X}
  (v,(P\otimes P')v)>(1-\epsilon) (v,(P\otimes I)v),
  \end{equation}
 while, on the other,
 \begin{equation}
 \label{eq:Y}
 (v,(P\otimes P')v)=(v,(P\otimes I)v)(v,(I\otimes
 P')v)
 \end{equation}
 since $v$ is a product state with respect to $H = H_{S} \otimes
 H_{S'}$.  Taken together, Eqns. \ref{eq:X} and \ref{eq:Y} entail $(v,(I\otimes
 P')v)>1-\epsilon$.  But since $\epsilon$ can be chosen arbitrarily small,
 $(v,(I\otimes P')v)=1$ for any nonzero
 projection $P'\in S'$, which of course is absurd.

 \section{Tests for Hyperentanglement}

 In order to be able to say more about hyperentanglement, we turn now
 to establishing the equivalence of $S$-cyclicity and maximal
 correlation of algebras with two further conditions ($3.$ and $4.$
 below) which serve
 as simple tests of a state's hyperentanglement.

\begin{theorem}
\label{equivalents}
Let $v$ be any state vector in $H = H_{S} \otimes
 H_{S'}$, $D_{S'}^{v}$ be
the reduced density operator for the subsystem $S'$ as determined by
$v$, and $\{b_{j}\}$ be any orthonormal basis
for $H_{S'}$.  Then the following are equivalent:
\begin{enumerate}
\item $v$ is $S$-cyclic.
\item $S'$ is maximally correlated with $S$ in state $v$.
\item $D_{S'}^{v}$ does not have $0$ as an eigenvalue.
\item $v$ may be expanded as $v = \sum_{j}
v_{j}  \otimes  b_{j}$ where the set of vectors $\{v_{j}\}\subseteq H_{S}$
is linearly independent.
  \end{enumerate}
\end{theorem}
\emph{Proof}.  $1.\Rightarrow 2.$  This is just Redhead's theorem
(Thm. 1 above).

$2.\Rightarrow 3.$  Let $w$ be any nonzero vector in
$H_{S'}$ and $P_{w}'$ be the projection onto the subspace generated by $w$.
Clearly a necessary condition for $S'$ to be maximally correlated with
$S$ in state $v$ is that
Prob$_{v}(P_{w}'=1)>0$.  It follows that
$Tr(D_{S'}^{v}P_{w}')\not
=0$, and hence that $D_{S'}^{v}w\not
=0$.  Since $w$ was arbitrary, $D_{S'}^{v}$ cannot have $0$ as an
eigenvalue.

$\neg 4.\Rightarrow \neg 3.$  Since we can pick an
orthonormal basis $\{a_{i}\}\subseteq H_{S}$ and expand any vector in $H$,
in particular $v$, as
\begin{equation}
v=\sum_{ij}c_{ij}
a_{i}  \otimes  b_{j}=\sum_{j}(\sum_{i}c_{ij}a_{i})\otimes b_{j},
\end{equation}
there exist vectors $\{v_{j}\}\subseteq H_{S}$ such that $v = \sum_{j}
v_{j}  \otimes  b_{j}$.  Assuming $\neg 4.$ (i.e. $4.$ is false), at least
one element of $\{v_{j}\}$ must be
a finite linear combination of other elements in $\{v_{j}\}$.  (To
avoid possible confusion, note that we are using the general notion of linear
independence, applicable to any vector space (topological or not), according
to which a set of vectors is linearly independent exactly when all its
finite subsets are.)  Without
loss of generality, we may assume that $v_{1}$ is a linear combination
of $\{v_{2},\ldots,v_{m}\}$---for if this were not the case, we could always
renumber the basis vectors
$\{b_{j}\}$ so that it is.  Thus, for some coefficients
$\{c_{j}\}_{j=2}^{m}$
we have
\begin{equation}
v_{1}=\sum_{j=2}^{m}c_{j}v_{j}.
\end{equation}
This gives
\begin{eqnarray}
v & = & v_{1}  \otimes  b_{1}+\sum_{j>1}
v_{j}  \otimes  b_{j} \\
& = & (\sum_{j=2}^{m}c_{j}v_{j})  \otimes  b_{1}+\sum_{j>1}
v_{j}  \otimes  b_{j} \\
\label{eq:zero}
& = & \sum_{j=2}^{m}v_{j}\otimes(c_{j}b_{1}+b_{j})+\sum_{j>m}
v_{j}  \otimes  b_{j}.
\end{eqnarray}
Now consider the subspace $T$ in $H_{S'}$ spanned by the
orthonormal vectors
$\{b_{j}\}_{j=1}^{m}\subseteq \{b_{j}\}$.  Clearly $T$ has dimension $m$,
and the $m-1$ vectors
$\{c_{j}b_{1}+b_{j}\}_{j=2}^{m}$ span a proper subspace of $T$.
Therefore there is at least one nonzero vector $w\in T$ orthogonal to all the
vectors $\{c_{j}b_{1}+b_{j}\}_{j=2}^{m}$.  Moreover, since $w\in T$,
$w$ is orthogonal to all the remaining basis vectors $\{b_{j}\}_{j>m}$ as
well.
So if we let $P_{w}'$ be the projection operator whose range is the
subspace generated by $w$, then the action of
$I\otimes P_{w}'$  on $v$---considering the expansion of $v$ given
 in Eqn. \ref{eq:zero}---produces the 0 vector.  It follows
that $Tr(D_{S'}^{v}P_{w}')=0$, and hence that $w$ is an eigenvector of
$D_{S'}^{v}$ corresponding to eigenvalue 0, so that condition 3. fails.

$4.\Rightarrow 1.$  As we saw at the beginning of the previous
argument, any vector $w\in H$ can be expanded as
$w = \sum_{j} w_{j}   \otimes  b_{j}$ where $\{w_{j}\}$ is some set
of vectors in $H_{S}$.  To establish $S$-cyclicity of $v$,
all we need to do is construct a sequence of
bounded operators
$\{A_{m}  \otimes  I\}$ which act on
$v  = \sum_{j} v_{j}  \otimes  b_{j}$ to bring it arbitrarily
close in norm to $w= \sum_{j} w_{j}   \otimes  b_{j}$.
So define the mapping $A_{m}$ ($m=1, 2, ...$)
by:
\begin{eqnarray}
A_{m} v_{j} & = &  w_{j}\ \mbox{if}\ j\leq m,\\
& = & 0\ \mbox{if}\ j> m.
\end{eqnarray}
Since the set $\{v_{j}\}$ is linearly independent by hypothesis, this
definition
extends to a linear operator $A_{m}$ from the closed subspace generated
by $\{v_{j}\}$ to $H_{S}$, an operator which is bounded, since (for finite $m$)
$A_{m}$ annihilates all but a finite number of the $\{v_{j}\}$.  If we
further extend the definition of $A_{m}$ to the whole of
$H_{S}$ by stipulating that $A_{m}u=0$ for all $u$ orthogonal to the
closed subspace generated by $\{v_{j}\}$, we get a bounded linear operator
$A_{m}$ acting on $H_{S}$ for each $m$.
Now just note that, by construction of the $\{A_{m}\}$,
$\lim (A_{m} \otimes I)v = w$.\ \emph{QED.}\vspace{.15in}

Looking at Bohm's \cite{bohm} version of the EPR state using
two spin-$1/2$ particles
\begin{equation}
  \frac{1}{\sqrt{2}}[|\sigma_{1z}=+1\rangle|\sigma_{2z}=-1\rangle+
  |\sigma_{1z}=-1\rangle|\sigma_{2z}=+1\rangle],
  \end{equation}
we see straightaway, using condition Thm. $2.4$ (i.e.
condition $4.$ of Thm. $2$), that $1$ is maximally
correlated with $2$ and vice-versa.  In fact, \emph{every} entangled state
of two spin-$1/2$ particles is hyperentangled.  For it is easy to
verify that a state $v\in H_{1}^{2}\otimes H_{2}^{2}$ (here, the
superscripts indicate dimension) is entangled if
and only if neither $D_{1}^{v}$ nor $D_{2}^{v}$ have $0$ as an
eigenvalue which, in view of Thm. $2.3$, is
exactly the condition for hyperentanglement.  Thus the entangled spin-1/2
Hardy state \cite{hard,clif}
\begin{equation}
\frac{1}{\sqrt{3}}[2|\sigma_{1x}=+1\rangle|\sigma_{2x}=+1\rangle-
  |\sigma_{1z}=+1\rangle|\sigma_{2z}=+1\rangle],
\end{equation}
is hyperentangled too.

This conclusion that every entangled state is hyperentangled is
actually an artifact of $H$ having two-dimensional factors.
For the same reason that Bohm's singlet state is
hyperentangled, the spin-1 singlet state
\begin{equation}
  \frac{1}{\sqrt{3}}[|S_{1y}=0\rangle|S_{2y}=0\rangle
  -|S_{1x}=0\rangle|S_{2x}=0\rangle
  -|S_{1z}=0\rangle|S_{2z}=0\rangle]
 \end{equation}
employed by Heywood and Redhead \cite{hr} in their algebraic proof
  of nonlocality is hyperentangled.
   But the closely related entangled state
  \begin{equation}
  |v\rangle = \frac{1}{\sqrt{2}}[|S_{1y}=0\rangle|S_{2y}=0\rangle
  -|S_{1x}=0\rangle|S_{2x}=0\rangle]
  \end{equation}
  is not, since relative to the orthonormal basis
  \begin{equation}
   b_{1}=|S_{2y}=0\rangle,\ b_{2}=|S_{2x}=0\rangle,\
   b_{3}=|S_{2z}=0\rangle,
   \end{equation}
   for $H_{2}^{3}$, the corresponding
  vectors
  \begin{equation}
  v_{1}=\frac{1}{\sqrt{2}}|S_{1y}=0\rangle,\
v_{2}=-\frac{1}{\sqrt{2}}|S_{1x}=0\rangle,\ v_{3}=|0\rangle,
  \end{equation}
  are obviously not linearly independent, in violation of Thm. $2.4$
  for $S=1$. (A similar violation occurs for $S=2$, but of course the
  $S=1$ violation is enough to defeat the hyperentanglement.)

   When $H$ has more than two factors, the possibility of
  hyperentanglement looks even less likely.  The three spin-1/2
  Greenberger-Horne-Zeilinger \cite{ghz} state
  \begin{equation}
  \frac{1}{\sqrt{2}}[|\sigma_{1z}=+1\rangle
  |\sigma_{2z}=+1\rangle|\sigma_{3z}=+1\rangle+
  |\sigma_{1z}=-1\rangle|\sigma_{2z}=-1\rangle|\sigma_{3z}=-1\rangle]
  \end{equation}
  fails the test of Thm. 2.3 in the case $S=1$, because all vectors 
  orthogonal to both $|\sigma_{2z}=+1\rangle|\sigma_{3z}=+1\rangle$
  and $|\sigma_{2z}=-1\rangle|\sigma_{3z}=-1\rangle$ in
  $H_{2}^{2}\otimes H_{3}^{2}$ are eigenvectors
  of the density operator for $\{2,3\}$ with eigenvalue $0$ (similar remarks
  applying for the density operators of $\{1,2\}$ and $\{1,3\}$).
  We leave the reader the task of checking that
  the three spin-1/2 Hardy state \cite{clif}
\begin{equation}
\frac{1}{\sqrt{7}}[2^{3/2}|\sigma_{1x}=+1\rangle|\sigma_{2x}=+1\rangle|\sigma_{3x}=+1\rangle-
  |\sigma_{1z}=+1\rangle|\sigma_{2z}=+1\rangle|\sigma_{3z}=+1\rangle],
\end{equation}
also fails to pass muster.

These last two nonexamples suggest that hyperentangled states may not be
all that
common in quantum theories after all.  On the contrary,
we shall now show that when a compound system actually has a state space
that permits the existence of hyperentangled states, they must be
norm dense in the unit sphere of that space.

\section{The Existence and Density of Hyperentangled States}

We start by noting that the test of hyperentanglement supplied by the
conditions of Thm. 2 can be simplified so that it is only necessary to
check satisfaction of (any one of) those conditions for the `atomic'
subsystems represented by the individual factors $H_{1},
H_{2},\ldots,H_{n}$ of $H$.

\begin{theorem}
A state $v\in H$ is hyperentangled if and only if any (and therefore
all) of Thm. 2's equivalent conditions on $v$ hold in all the cases
$S=1,2,\ldots,n$.
\end{theorem}
 \emph{Proof}.  `Only if'.  If $v$ is hyperentangled, then  (by definition)
for any
 two nonoverlapping subsystems $S$ and $T$, the algebra $S$ is
 maximally correlated to the algebra $T$ in state $v$.  In
 particular, for any index $i$ the algebra $i'$ (corresponding to the set
 of indices in $\{1,\ldots,n\}$ unequal to $i$) is maximally correlated to
the algebra
 $i$, so that Thm. $2.2$ holds for $v$ in the case $S=i$.

 `If'.  Consider any two nonoverlapping subsystems $S$ and $T$.  If
 $j$ is any index in the set $T$, then since by hypothesis Thm. 2
 holds for $v$ when $S=j$, $v$ is $j$-cyclic which \emph{ipso facto}
 means $v$ must be $T$-cyclic.  Using
 $1.\Rightarrow 2.$ of Thm. 2, this entails that $T'$ is maximally
 correlated with $T$ in state $v$, so that in particular $S$ is maximally
 correlated with $T$ in state $v$, because $S\subseteq T'$.\ \emph{QED.}\vspace{.15in}

 For our next result below, we shall be employing Thm. $3$ and in
 particular the equivalence between hyperentanglement, Thm. $2.3$ for
 $S=1$ to $n$ (the condition on the atomic system's density operators),
 and Thm. $2.4$ for $S=1$ to $n$ (the linear independence condition).

 \begin{theorem}
\label{density}
Let the state space be given by $H= H_{1} \otimes H_{2} \otimes\cdots
\otimes H_{n}$ where $n>1$ and each factor space is separable and has
nontrivial ($>1$) dimension.  Then the following are equivalent:
\begin{enumerate}
\item There exists a hyperentangled state in $H$.
\item All the factors of $H$ have the same dimension, and if $n>2$ their
(common)
dimension is infinite.
\item The set of hyperentangled states is norm dense in the unit sphere
of $H$.
  \end{enumerate}
\end{theorem}
\emph{Proof}.  $1.\Rightarrow 2.$  We begin by establishing:

\begin{center}
 \textbf{Lemma}:
For $i=1$ to $n$, $\dim H_{i} = \dim H_{i'}$.
\end{center}

\noindent (As before, we write $i$ for $\{i\}$ and $i'$ for
$\{i\}'=\{j:j\not =i\}$.)

By hypothesis, some state $v\in H$ is hyperentangled.  Let
 \begin{equation}
 \label{eq:biorthogonal}
v = \sum_{k} c_{k}(a_{k} \otimes b_{k})
\end{equation}
be a Schmidt decomposition of $v$ with respect to the factorization
$H = H_{i}
 \otimes H_{i'}$, where $\{a_{k}\}$ and $\{b_{k}\}$ are orthonormal sets in
$H_{i}$ and
$H_{i'}$ respectively.  If
$\dim H_{i}
< \dim H_{i'}$, then the set $\{b_{k}: c_{k}\neq 0\}$ cannot form a basis for
$H_{i'}$.  Therefore there must be a nonzero vector $w\in H_{i'}$
orthogonal to all the
vectors in
$\{b_{k}: c_{k}\neq 0\}$.  If $P'_{w}$ is the projection onto the
subspace $w$ generates, then by Eqn. \ref{eq:biorthogonal} $(I\otimes
P'_{w})v=0$, and $w$ must be an eigenvalue of $D_{i'}^{v}$
corresponding to eigenvalue $0$.  But since $v$ is hyperentangled,
this contradicts Thm. $2.3$ for $S=i$.

All that remains is to show that $\dim H_{i} > \dim H_{i'}$ leads
to a
similar contradiction, and then the \textbf{Lemma} follows.
To this end, let $H_{j}$ be any one of the Hilbert space factors of $H$
that occurs in
$H_{i'}$
and re-express $H$ as $H = H_{j}  \otimes  H_{j'}$.
Since
\begin{equation}
\label{eq:helloAlex}
\dim H_{j}\leq \dim H_{i'} < \dim H_{i} \leq \dim H_{j'},
\end{equation}
$\dim H_{j} < \dim H_{j'}$.
So we are in exactly in the same situation as we were before; that is,
by our hypothesis that
there is a hyperentangled $v\in H$, $v$ must (in particular) satisfy
Thm. $2.3$ for $S=j$, and we can run through the argument of the previous
paragraph, with $j$ in place of $i$, to get a contradiction with
$\dim H_{j} < \dim H_{j'}$.

With \textbf{Lemma} in hand, the proof that $1.\Rightarrow 2.$ is
now straightforward.  If n=2, then the \textbf{Lemma} immediately yields that
both factors of
$H$
must have the same dimension.  If $n>2$, write $H$ as
$H= H_{i}  \otimes  H_{j}  \otimes  H_{\{i,j\}'}$
isolating the $i$th and $j$th Hilbert space factors in $H$ and denoting the
tensor
product of the rest of the factors by
$H_{\{i,j\}'}$.  Again using the \textbf{Lemma}:
\begin{equation}
\label{eq:A}
\dim H_{i}  =  \dim [ H_{j}  \otimes  H_{\{i,j\}'}] = \dim H_{j} \dim
H_{\{i,j\}'},
\end{equation}
\begin{equation}
\label{eq:B}
\dim H_{j}  =  \dim [ H_{i}  \otimes  H_{\{i,j\}'}] = \dim H_{i} \dim
H_{\{i,j\}'}.
\end{equation}
Since $\dim H_{\{i,j\}'} > 1$,
there is no solution to Eqns. \ref{eq:A} and
\ref{eq:B} when
either
$\dim H_{i}$ or $\dim H_{j}$ is finite.  Since $i$ and $j$ were
arbitrary, this shows that
\emph{all} of $H$'s factors must be infinite-dimensional.

$2.\Rightarrow 3.$  Observe first that since the Hilbert space $H$
is a complete metric space (by definition),
its unit
sphere is closed and defines a complete metric
subspace of $H$.  Now suppose we could establish the following:

\begin{center}
\textbf{Claim}: For any
$i=1$ to $n$, the set of states satisfying Thm. $2.4$ for $S=i$ is a
countable intersection of dense open sets in the unit sphere of $H$.
\end{center}

\noindent If so, then since hyperentanglement amounts to
satisfying Thm. $2.4$ for all $S=1$ to $n$ (cf. Thm.
$3$), the set of all hyperentangled states
would \emph{also} have to be a countable
intersection of dense open sets in the unit sphere of $H$.
But the Baire category
theorem \cite{BCT} asserts
that in a complete metric space---such as the unit sphere of $H$---a
countable
intersection of dense open sets must itself be dense!  So we would
have the desired conclusion if we could establish the \textbf{Claim},
which we now proceed to do.

Fix, once and for all, an $i$ such that $1\leq i\leq n$.  Note that, by
hypothesis, all of $H$'s factors have the same dimension,
and infinite dimension when $n>2$.  So in either case we have
$\dim H_{i} = \dim H_{i'}$.  Also fix (once and for all) an orthonormal basis
$\{b_{j} \}$ for $H_{i'}$.  Let $T$ be any
finite
subset of the indices that enumerate the vectors $\{b_{j} \}$.
Remembering that each $v \in H_{i}  \otimes  H_{i'}$
can be expanded (indeed uniquely, given $\{b_{j} \}$) as
$v = \sum_{j} v_{j}  \otimes  b_{j}$, define  the set of states:
\begin{equation}
\label{eq:F}
F(T) = \{v \in H_{i}  \otimes  H_{i'}: \|v\|=1\ \mbox{and}\ \{v_{j}\}_{j\in
T}\
\mbox{is linearly independent}\}.
\end{equation}
Clearly a state $v\in H_{i}  \otimes  H_{i'}$ satisfies Thm. $2.4$
for $S=i$
if and only if, for every finite subset $T$ of the indices
enumerating $\{b_{j} \}$, $v\in F(T)$.
Since there are at most countably many
such
\emph{finite} subsets (even if the basis $\{b_{j} \}$ is infinite), it
suffices for
the \textbf{Claim} to show that for any finite $T$,
$F(T)$ is both
dense
and open in the unit sphere of $H$.

1) \emph{$F(T)$ is norm dense in the unit sphere of $H$}.  Choose any state
 $w \in H_{i}  \otimes  H_{i}'$ and
let
\begin{equation}
\label{eq:see}
w = \sum_{k \in K} c_{k}(x_{k}  \otimes  y_{k})
\end{equation}
be a Schmidt decomposition for $w$.  Here the $\{c_{k}\}$ are
coefficients (not necessarily all nonzero),
$\{x_{k}\}$ and $\{y_{k}\}$ are orthonormal bases in
$H_{i}$ and $H_{i'}$
respectively, and $K$ is either a finite or countably infinite index
set (depending on whether $H_{i}$ and $H_{i'}$ are both finite or
infinite-dimensional).  For density, we need to show that we can always find a
state in $F(T)$ arbitrarily close to $w$.  There are two cases.

(Case 1): $c_{k}$ is nonzero for all $k\in K$.  In this case, $w$ is
itself in
$F(T)$.   For the expansion of $w$ in Eqn. \ref{eq:see} satisfies
Thm. $2.4$ for $S=i$, taking the orthonormal basis of $H_{i'}$ in that
theorem
to be $\{y_{k}\}$.  However, the equivalences in Thm. $2$ hold
no matter what orthonormal basis
for $H_{i'}$ we choose, so that satisfaction of Thm. $2.4$ for $S=i$
relative to one such
basis entails satisfaction relative to them all.  Thus
relative to the specific orthonormal basis $\{b_{j}\}$ in the definition of
$F(T)$, $w=\sum_{j}w_{j}\otimes b_{j}$ must be such that the
vectors $\{w_{j}\}\subseteq H_{i}$ are linearly independent, and in
particular $w\in F(T)$.

(Case 2): $c_{k}$ is zero for at least one $k\in K$.  If so, consider
the family of
states of the form:
\begin{equation}
u = \frac{\sum_{k \in K} d_k (x_{k}  \otimes  y_{k})}{\sum_{k \in K}
|d_k|^{2}}
\end{equation}
where $d_{k}=c_{k}$ if $c_{k}\not = 0$, $d_{k}\not =0$ if $c_{k}= 0$,
and the sequence $\{d_{k}:c_{k}=0\}_{k\in K}$ is square-summable.
(Note that since $\dim H_{i} = \dim H_{i'}$, there are indeed states
in $H$ of form $u$.) By the argument of (Case 1), every state of form $u$
lies
in $F(T)$.  Moreover, we can make $u$ as close to $w$ as we like by making
the
coefficients $d_{k}$ corresponding to $c_{k} = 0$ arbitrarily small.

2) \emph{$F(T)$ is open in the unit sphere of $H$}.
Let $B$ denote the span in $H_{i'}$ of $\{b_{j} \}_{j \in T}$.
If $W$ is any subspace of $H_{i}$ of (finite) dimension $m = \dim(B) =
|T|$,
the projection $P_{W} \otimes P_{B}$ maps $H$ onto $W \otimes B$.
Applied to $v = \sum_{j} v_{j} \otimes b_{j} \in H$, this projection yields
$\sum_{j \in T} P_{W}v_{j} \otimes b_{j}$.
Let $F(W,T)$ denote the
        set of those vectors $v = \sum_{j} v_{j} \otimes b_{j} \in H$
for which
        $\{P_{W}v_{j}\}_{j\in T}$ is linearly independent.
        This last implies that $\{v_{j}\}_{j \in T}$ is
linearly independent as well, so we have $F(W,T)\cap X \subseteq
F(T)$ where $X$ is the unit
sphere of $H$. On the other
hand, if $v\in F(T)$, then taking $V =  \mbox{span}\{v_{j}\}_{j\in T}$,
plainly $v\in F(V,T)$.
        Thus, $F(T) = \bigcup_{W} F(W,T) \cap X$
        where the union is taken over all $m$-dimensional subspaces of
$H_{i}$.

        To show that $F(T)$ is open, it now suffices to show that each
$F(W,T)$ is
        open.  Let $U$ denote the collection of vectors $u \in W \otimes
B$ of the
        form $\sum_{j\in T} u_{j} \otimes b_{j}$ with
        $\{u_{j}\}_{j\in T}$ linearly
independent.
        Since $F(W,T) = (P_{W} \otimes P_{B})^{-1}(U)$ and
        projections are continuous, our task further
reduces
        to that of showing that $U$ is open in $W \otimes B$.

        Choosing a basis $\{w_{i}\}_{i \in T}$ for $W$, each vector $u
\in W \otimes
        B$ has a unique expansion as $u = \sum_{i,j \in T} c_{ij} w_{i}
\otimes
        b_{j}$. Let $[u] = [c_{ij}]$ denote the $m \times m$ matrix
consisting of
        the coefficients $c_{ij}$, and define $\det : W \otimes B
\rightarrow C$ by $\det(u) := \det[u]$. Note that $\det$ is continuous
(and in fact
        independent of the choice of $\{w_{i}\}_{i \in T}$). Notice also
that if we
express $u \in
        W \otimes B$ as $\sum_{j\in T} u_{j} \otimes b_{j}$, where $u_{j} =
\sum_{i\in T}
        c_{ij} w_{i}$, then since $\det(u) = 0$ just in case the columns of
        $[c_{ij}]$ are linearly dependent, $\det(u)=0$ exactly when the vectors
        $\{u_{j}\}_{j\in T}$
are linearly
        dependent.  It follows that $U = \det^{-1}(C \setminus
        \{0\})$,
and since $\det$ is
        continuous, $U$ is open in $W \otimes B$ as claimed.\ \emph{QED.}\vspace{.15in}

 The above theorem lays bare the fundamental obstacle to
 states of three or more spin-1/2 particles being hyperentangled:
 hyperentangled states cannot live in finite-dimensional state spaces
 when there are more than two particles!  (In fact, this was
 anticipated by Wagner \cite{wag}, but his arguments do not
 establish it in full generality.)

 Moreover, the density of hyperentangled states (when they
 exist) entails that such states need not differ from other
 entangled states simply by their degree of entanglement.
 Recently Shimony \cite{shim} has proposed the following definition
 for
 the
 degree of entanglement of a state $v\in H$:
 \begin{equation}
 E(v)=\frac{1}{2}\mbox{min}\|v-w\|^{2}
 \end{equation}
 where $w$ is a product state in $H$ and the minimum is taken over the
 set of all product states.  Shimony appears to propose this only for the case
 $n=2$ but there is no reason not to adopt his definition in general, so
that the
 minimum is taken over all $n$-fold product states.  Shimony shows
 that if $H$ is finite-dimensional $E(v)\in (0,1]$ ($E(v)=0$
 corresponding to no entanglement), whereas in the infinite-dimensional
 case no state actually attains an entanglement degree of $1$.  In any
 case, since hyperentangled states are dense in $H$ (when its
 factors have the appropriate dimension),
 it follows that their degrees of
 entanglement lie dense in the
 interval $(0,1)$, and in particular, that there are hyperentangled states
 with degrees of entanglement arbitrarily close to 0!
 This shows that hyperentanglement, despite the pervasive maximal
 correlations involved, should be viewed as a new
 \emph{kind} of entanglement, and certainly not as a case of
 maximal \emph{entanglement}.

  \section{Constructing Hyperentangled States}

An unusual feature of our proof that hyperentangled states are
dense
is that in the most interesting case of $n>2$, when the factor
spaces of $H$ must be infinite-dimensional, our proof is not
constructive because it relies essentially on the Baire category
theorem.  Popescu has conjectured (\cite{wag}, p. 32), by
analogy with the vacuum state,
that the ground state of $n$ quantized coupled harmonic oscillators should
provide an explicit example of (what we have been calling) a
hyperentangled state.  But while Wagner (\cite{wag}, pp. 32-4) has
confirmed this
for $n=2$, our own efforts to find suitable couplings
in the case $n>2$ have not been successful.  This in turn raises the
more general question of
 what conditions on the Hamiltonian governing a collection of
 $n>2$ systems will
 guarantee that it will spend most of its time in a hyperentangled
 state.

 On the other hand, with a little ingenuity it is perfectly possible to
 write down a
state that is hyperentangled when $n>2$.   Expand a general state $w\in H$ in
terms of a product basis for $H=H_{1}\otimes H_{2}\otimes\cdots\otimes
H_{n}$ as
\begin{equation}
w=\sum_{a,b,\ldots,z\in N}v_{a,b,\ldots,z
}y^{1}_{a}\otimes y^{2}_{b}\otimes\cdots \otimes y^{n}_{z}
\end{equation}
where $N$ is the natural numbers (including $0$) and $[v_{a,b,\ldots
z}]$ are the elements of $w$'s square-summable and countable by countable by
countable\ldots ($n$ times)
coefficient matrix.
Using Thm. $2.4$ for $S=1,\ldots,n$ it is easy to see
that $w$ will be hyperentangled exactly when the rows of $[v_{a,b,\ldots
z}]$
are linearly independent and the columns of $[v_{a,b,\ldots
z}]$ are
linearly independent and the `files' of $[v_{a,b,\ldots
z}]$ are linearly independent, etc.

Of course, it is trivial to find a square-summable coefficient matrix of
this sort when $n=2$.
For $n>2$, things get more complicated. We shall give two methods for
constructing a suitable coefficient matrix in this case in the hope that
one of them might be
parlayed into an actual \emph{physical} example of a hyperentangled
state.

For the first method, start with the case $n=3$.
Fix an injection $j:N\times N\rightarrow N$ such that $j(a,b) \geq
max(a,b)$.  (For example, we could choose $j(a,b)=2^{a}3^{b}$.)
Consider those functions $h:N\times N\times N\rightarrow C$ such that
\begin{equation}
 h(a,b,c)\not=0\Leftrightarrow \mbox{either}\ a=j(b,c)\ \mbox{or}\
 b=j(a,c)\ \mbox{or}\ c=j(a,b).
 \end{equation}
We shall need to employ the following preliminary result.
\begin{center}
 \textbf{Lemma}:
$c>j(a,b)$ (resp. $a>j(b,c)$, resp. $b>j(a,c)$) implies $h(a,b,c)=0$.
\end{center}
\emph{Proof}.  Suppose that $c>j(a,b)$ and $h(a,b,c)\not=0$.  Then
$c\not=j(a,b)$, so either
$a=j(b,c)$ or $b=j(a,c)$.  But if $a=j(b,c)$, then
 $c>j(j(b,c),b) \geq max(j(b,c),b) \geq max(max(b,c),b)=max(b,c)$
a contradiction since obviously $c \leq max(b,c)$.  Similarly, if
$b=j(a,c)$, then
 $c>j(a,j(a,c)) \geq max(a,j(a,c)) \geq max(a,max(a,c))=max(a,c)$
again a contradiction.  So $h(a,b,c)=0$, as desired.
(By symmetry, the same conclusion follows if
$a>j(b,c)$ or $b>j(a,c)$.)\ \emph{QED.}\vspace{.15in}

We now write $v_{a,b,\cdot}$ for the row vector $(h(a,b,c))_{c \in N}$
(and similarly $v_{\cdot,b,c}$ for the column vector $(h(a,b,c))_{a \in N}$
and $v_{a,\cdot,c}$ for the `file' vector $(h(a,b,c))_{b \in N}$).

\begin{theorem}
$\{v_{a,b,\cdot}\}_{a,b\in N}$
(resp. $\{v_{\cdot,b,c}\}_{b,c\in N}$, resp.
$\{v_{a,\cdot,c}\}_{a,c\in N}$)
constitutes a linearly independent set.
\end{theorem}
\emph{Proof}.  Suppose (contrary to hypothesis) that
 $\sum_{i=1}^{m} s_i v_{a_i,b_i,\cdot} = 0$ with all coefficients $s_i\not=0$
and all pairs $(a_i,b_i)$ distinct.  Then for all $k\in N$,
$\sum_{i=1}^{m} s_i h(a_i,b_i,k)=0$.  In particular, setting
$k'=max_{i=1}^{m} j(a_i,b_i)$ we have
\begin{equation}
\label{eq:David}
\sum_{i=1}^{m} s_i h(a_i,b_i,k')=0.
\end{equation}
But by the distinctness of the $(a_i,b_i)$, the injectivity of the $j$
function
and the \textbf{Lemma} above,
the sum on the left-hand side of Eqn. \ref{eq:David}
involves exactly one non-zero term,
a contradiction which proves the linear independence of
$\{v_{a,b,\cdot}\}_{a,b\in N}$.
  (The linear independence of each of the sets
$\{v_{\cdot,b,c}\}_{b,c\in N}$ and
$\{v_{a,\cdot,c}\}_{a,c\in N}$ follows by symmetry.).\ \emph{QED.}\vspace{.15in}

This little theorem supplies a recipe for building infinite 3-dimensional
matrices with
linearly independent rows,  linearly independent columns, and
linearly independent files.   Such matrices must have non-zero
values at position $(a,b,c)$ exactly when
either $a=j(b,c)$, $b=j(a,c)$ or $c=j(a,b)$, but then we have the freedom
to choose as we please any non-zero value for $h(a,b,c)$.  In particular,
we can easily arrange for the square-summability of the entries of the
matrix $[v_{a,b,c}]$.

(For a {\em bijection}
$j:N\times N\rightarrow N$ such that $j(a,b) \geq max(a,b)$,
$\{v_{a,b,\cdot}\}_{a,b\in N}$
(resp. $\{v_{\cdot,b,c}\}_{b,c\in N}$, resp.
$\{v_{a,\cdot,c}\}_{a,c\in N}$) actually constitutes a Hilbert
space basis.  Indeed the set of vectors with finitely many non-zero
entries lies dense in Hilbert space and we can express any such vector
as a linear combination of vectors
from
$\{v_{a,b,\cdot}\}_{a,b\in N}$
(resp. $\{v_{\cdot,b,c}\}_{b,c\in N}$, resp.
$\{v_{a,\cdot,c}\}_{a,c\in N}$).  Thus, consider any
vector $(q_i)_{i \in N}$ with last non-zero entry $q_I$.
Then the vector $(q_i) - (q_I/h(j^{-1}(I),I)) v_{j^{-1}(I),\cdot}$ has its
last non-zero entry before position $I$, and we may proceed
inductively.  For an explicit example of a bijective $j$ function, define
$j(a,b)$ so it has binary expansion
$\beta_n\alpha_n \ldots \beta_1\alpha_1$ when $a$ has binary
expansion $\alpha_n \ldots \alpha_1$ and $b$ has binary
expansion $\beta_n \ldots \beta_1$.)

The arguments above extend {\em mutatis mutandi} to provide
higher dimensional matrices with the analogous property.  To get a
4-dimensional matrix, for example, we should consider functions $h$ where
$h(a,b,c,d)\not=0$ if and only if either
$a=j(b,j(c,d))$, $b=j(a,j(c,d))$, $c=j(a,j(b,d))$ or $d=j(a,j(b,c))$.
We may even construct analogous
infinite-dimensional matrices of the required sort, with entries
indexed by finitely non-zero natural number sequences.

We turn now to describing a second independent method of constructing
coefficient matrices
for hyperentangled states.  Let $v$ denote any $p\times p\times p$-matrix with
complex entries, viewing $v$ as a function $v:\{1,\ldots,p\}^3\rightarrow C$.
Write $v_{a,b,\cdot}$ for the vector $(v(a,b,c))_{c=1}^{p}$
(and similarly $v_{\cdot,b,c}$ for the vector $(v(a,b,c))_{a=1}^{p}$
and $v_{a,\cdot,c}$ for the vector $(v(a,b,c))_{b=1}^{p}$).
Ultimately we shall
use such a matrix $v$ as the `seed' from which to a `grow' a countable
by countable by countable matrix with the required properties.  Again,
we need a preliminary result to make this construction go through.

\begin{theorem}
Let $v$ denote a $p\times p\times p$ matrix with complex entries.
Fix $m\leq p$ and assume the linear independence of the vector sets
$\{ v_{a,b,\cdot} \}_{1\leq a,b \leq m}$,
$\{ v_{a,\cdot,c} \}_{1\leq a,c \leq m}$ and
$\{ v_{\cdot,b,c} \}_{1\leq b,c \leq m}$.
Now set $p'=p^2+p-m^2$.
Then the matrix $v$ has an extension to a $p'\times p'\times p'$-matrix $v'$
with
\begin{enumerate}
\item linearly independent vector sets 
$\{ v'_{a,b,\cdot} \}_{1\leq a,b \leq p}$,
$\{ v'_{a,\cdot,c} \}_{1\leq a,c \leq p}$ and \newline
$\{ v'_{\cdot,b,c} \}_{1\leq b,c \leq p}$;
\item $v'(a,b,c)=0$ whenever
$a,b\leq m$ and $c>m$, $a,c\leq m$ and $b>m$ or $b,c\leq m$ and $a>m$.
\end{enumerate}
Moreover we can arrange to have the sum of the squares of the
absolute values of the new entries equal to any
$\epsilon>0$.
\end{theorem}
\emph{Proof}.
First we arrange for the vectors in
 $\{ v'_{a,b,\cdot} \}_{1\leq a,b\leq p}$
to extend the corresponding vectors in
 $\{ v_{a,b,\cdot} \}_{1\leq a,b \leq p}$.
For the sake of condition 2. we must certainly stipulate that
$v'(a,b,c)=0$ whenever $a,b\leq m$ and $c>m$.
This leaves us  with $p^2 - m^2$ length $p$ vectors to extend, namely
\begin{equation}
\{ v_{a,b,\cdot}: 1\leq a,b \leq p\ \mbox{and}\ a,b\ \mbox{are not both}\leq
m\},
\end{equation}
and we need to extend each of these to a length $p+p^2-m^2$ vector.  
Thus we
must append $p^2-m^2$ entries
to each vector.  Do this by simply appending
the $p^2-m^2$ standard basis
vectors in $C^{p^2-m^2}$ 
multiplied by
$\sqrt{\epsilon/(3(p^2-m^2))}$ 
(in any order).  
This makes the vector set
$\{ v'_{a,b,\cdot} \}_{1\leq a,b \leq p}$ linearly independent
because we could project  a linear dependence 
to a linear dependence either among 
the standard basis vectors in $C^{p^2-m^2}$ or among the vectors
$\{ v_{a,b,\cdot} \}_{1\leq a,b \leq m}$.

To finish, we repeat the procedure so as to extend likewise the vectors
$\{ v_{a,\cdot,c} \}_{1\leq a,c \leq p}$ and
$\{ v_{\cdot,b,c} \}_{1\leq b,c \leq p}$.  Finally, we may set the values
of the remaining entries of $v'$ to $0$.\
\emph{QED.}\vspace{.15in}

Using this theorem, we can now build up the coefficient matrix for
an $n=3$
hyperentangled state as follows.  Begin, say, with $p=2$, $m=1$ and $v$ any
$2\times 2 \times 2$-matrix with $v(1,1,1)\not=0$.
Now iterate the process described in Thm. 6.  (Each
time the new $p$ equals the old $p'$ and the new $m$ equals the old $p$.) 
Observe that each row, column and file has only finitely many
nonzero entries, since it stabilizes at some finite stage of the
process.
Thus, we
obtain an infinite 3-dimensional matrix with linearly independent row
sets, column sets and file sets; 
any potential dependence involves just finitely many vectors, all
of these stable at some finite stage of the construction, so
a dependence would contradict the theorem.
\vspace{.15in}

\noindent \textbf{Acknowledgements}

 We would like to thank Don Hadwin (New Hampshire),
 John Norton (Pittsburgh), and David Malament
 (Chicago) for helpful discussions and comments.


\begin{thebibliography}{99}
\bibitem{but} J. N. Butterfield, in \emph{Fundamental Problems in 
Quantum Theory}, edited by D.M. 
Greenberger and A. Zeilinger (Annals of the New York Academy of 
Sciences \textbf{755}, 1995),  p. 768.
\bibitem{bell} J. S. Bell, \emph{Speakable and Unspeakable in 
Quantum Mechanics} (Cambridge University Press, Cambridge, 1987), Ch. 2.  
\bibitem{EPR}  A. Einstein, B. Podolsky, and N. Rosen, Phys. Rev. 
\textbf{47}, 777 (1935).
\bibitem{red1}  M. L. G. Redhead,  
Found. Phys. \textbf{25}, 123 (1995).   
\bibitem{red2}  M. L. G. Redhead,  in \emph{PSA 1994 Vol. 
2} (Philosophy of Science Association, East Lansing, MI, 
1995), p. 77.
\bibitem{wag}  F. Wagner, M. Phil. dissertation, Cambridge University (1997).
\bibitem{haag}  R. Haag, \emph{Local Quantum Physics}  
(Springer-Verlag, New York, 1992). 
\bibitem{horuzy}  S. Horuhzy, \emph{Introduction to Algebraic 
Quantum Field Theory} (Kluwer, Dordrecht, 1990).  
\bibitem{RS}  H. Reeh and S. Schlieder, Nuovo Cimento 
\textbf{22}, 1051 (1961). 
\bibitem{Schr}  E. Schr\"{o}dinger, Proc. Cam. Phil. Soc. \textbf{31}, 
555 (1935).  
\bibitem{fred}  K. Fredenhagen, J. Math. Phys. \textbf{7}, 1656 (1985). 
\bibitem{shim} A. Shimony, in \emph{Fundamental Problems in 
Quantum Theory}, edited by D.M. 
Greenberger and A. Zeilinger (Annals of the New York Academy of 
Sciences \textbf{755}, 1995), p. 675.
\bibitem{bohm}  D. Bohm, \emph{Quantum Theory}  
(Prentice-Hall, Englewood Cliffs, 1951) p. 611.
\bibitem{hard}  L. Hardy, Phys. Rev. Lett. \textbf{68}, 2981 (1992). 
\bibitem{clif}  C. Pagonis and R. Clifton, Phys. Lett. A
\textbf{168}, 100 (1992). 
\bibitem{hr} P. Heywood and M. L. G. Redhead,  
Found. Phys. \textbf{13}, 
481 (1983).
\bibitem{ghz}  D. M. Greenberger, M. A. Horne, and A. 
Zeilinger, in \emph{Bell's Theorem, Quantum Theory, and Conceptions of the 
Universe} edited by M. Kafatos (Kluwer, Dordrecht, 1989),  p. 69. 
\bibitem{BCT} W. A. Sutherland, \emph{Introduction to Metric and
Topological Spaces} (Oxford University Press, Oxford, 1975), p. 135.
\end{thebibliography}
\end{document}